\documentclass[12pt,preprint]{aastex}
\usepackage{graphicx}
\usepackage{natbib}

\shorttitle{High-Resolution Spectroscopy of TrES-1}
\shortauthors{Sozzetti et al.}

\begin{document}


\title{High-Resolution Spectroscopy of the Transiting Planet Host Star TrES-1}


\author{Alessandro Sozzetti\altaffilmark{1,2}, David Yong\altaffilmark{3},
Guillermo Torres\altaffilmark{2}, David
Charbonneau\altaffilmark{2}, David W. Latham\altaffilmark{2},
Carlos Allende Prieto\altaffilmark{4}, Timothy M.
Brown\altaffilmark{5}, Bruce W. Carney\altaffilmark{3}, and John
B. Laird\altaffilmark{6}} \altaffiltext{1}{Department of Physics
\& Astronomy, University of Pittsburgh, Pittsburgh, PA 15260 USA}
\altaffiltext{2}{Harvard-Smithsonian Center for Astrophysics, 60
Garden Street, Cambridge, MA 02138 USA}
\altaffiltext{3}{Department of Physics \& Astronomy,
University of North Carolina at Chapel Hill, Chapel Hill, NC 27599 USA}
\altaffiltext{4}{McDonald Observatory and Department of Astronomy,
University of Texas, Austin, TX 78712 USA}
\altaffiltext{5}{High Altitude Observatory/National Center for
Atmospheric Research, 3450 Mitchell Lane, Boulder, CO 80307 USA}
\altaffiltext{6}{Department of Physics \& Astronomy,
Bowling Green State University, Bowling Green, OH 43403 USA}
\email{asozzett@cfa.harvard.edu}
\email{yong@physics.unc.edu}
\email{gtorres@cfa.harvard.edu}
\email{dcharbon@cfa.harvard.edu}
\email{dlatham@cfa.harvard.edu}
\email{callende@hebe.as.utexas.edu}
\email{timbrown@hao.ucar.edu}
\email{bruce@physics.unc.edu}
\email{laird@tycho.bgsu.edu}


\begin{abstract}
We report on a spectroscopic determination of the stellar parameters and
chemical abundances for the parent star of the transiting planet
TrES-1. Based on a detailed analysis of iron lines in our Keck 
and HET spectra we derive $T_\mathrm{eff} = 5250\pm 75$ K,
$\log g = 4.6\pm 0.2$, and [Fe/H] $=  0.00\pm 0.09$. By measuring
the \ion{Ca}{2} activity indicator and by putting useful upper
limits on the Li abundance we constrain the age of TrES-1 to be
$2.5\pm 1.5$ Gyr. By comparing theoretical stellar evolution
models with the observational parameters we obtain $M_\star =
0.89\pm 0.05$ $M_\odot$, and $R_\star = 0.83\pm 0.05$ $R_\odot$.
Our improved estimates of the stellar parameters are utilized in a
new analysis of the transit photometry of TrES-1 to derive a mass
$M_p = 0.76\pm 0.05$ $M_\mathrm{J}$, a radius $R_p =
1.04^{+0.08}_{-0.05}$ $R_\mathrm{J}$, and an inclination $i =
89^\circ.5^{+0.5}_{-1.3}$. The improved planetary mass and radius
estimates provide the grounds for new crucial tests of theoretical
models of evolution and evaporation of irradiated extrasolar giant
planets.

\end{abstract}



\keywords{planetary systems: formation --- stars: individual
(TrES-1) --- stars: abundances}


\section{Introduction}

The discovery of an extra-solar Jupiter-sized planet transiting
the disk of the K0V star TrES-1 (Alonso et al. 2004) has marked
the first success of ground-based photometric surveys targeting
large areas of the sky with small-size telescopes searching for
low-amplitude periodic variations in the light curves of bright
stars (V $\leq 13$). In this Letter we report on a detailed
spectroscopic determination of the stellar parameters and chemical
abundances of iron and lithium of the transiting planet host star.

TrES-1 is only the second transiting planet orbiting a star bright
enough to allow for a variety of follow-up analyses similar to
those conducted for HD 209458b (see for example Charbonneau 2003,
and references therein).
In particular, high-resolution, high signal-to-noise ratio
spectroscopic observations of TrES-1 can be readily conducted,
allowing for improved values of the effective temperature, surface
gravity, and metallicity of the star. These parameters, when
compared with stellar evolution models, provide the means to
derive good estimates of its mass and radius. Finally, by
combining spectroscopic observations with transit photometry it is
possible to determine refined estimates of the mass and radius of
the planet. A better knowledge of these parameters offers the
tantalizing possibility for crucial tests of giant planet
formation, migration, and evolution. In addition, the hypothesis
of self-enrichment due to recent ingestion of planetary material
can be tested using detailed abundances of elements such as
lithium.

\section{Data Reduction and Abundance Analysis}

The spectroscopic observations which led to the radial-velocity
confirmation of the planetary nature of the transiting object
detected by the TrES (Trans-atlantic Exoplanet Survey) consortium
(Alonso et al. 2004) were performed with the HIRES spectrograph
and its I$_2$ absorption cell on the Keck I telescope (Vogt et al.
1994). Eight star+iodine spectra and one template spectrum were
collected over 18 days in July 2004. The template spectrum used here 
has a resolution $R\simeq 65,\,000$ and a signal-to-noise-ratio 
$S/N\simeq 80$ pixel$^{-1}$. Three $R\simeq 60,\,000$ spectra were taken
with the High Resolution Spectrograph (HRS, Tull 1998) on the
Hobby-Eberly Telescope (HET) during consecutive nights in August
2004, with spectral coverage in the range 5879-7838 \AA. The 
averaged spectrum has $S/N\simeq 120$ pixel$^{-1}$. For all spectra,
Thorium-Argon lamp exposures provided the wavelength
calibration.

Our abundance analysis of the Keck/HIRES and HET/HRS spectra of TrES-1
was carried out under the assumption of standard local
thermodynamic equilibrium (LTE) using a modified version of the
spectral synthesis code MOOG (Sneden 1973) and a grid of Kurucz
(1993) stellar atmospheres. We selected a set of 30 \ion{Fe}{1}
and 4 \ion{Fe}{2} lines (with lower excitation potentials
$0.86\leq\chi_l\leq 5.03$ eV and $2.58\leq\chi_l\leq 3.90$ eV,
respectively) from the HIRES spectrum, and used standard packages
in IRAF to derive equivalent widths (EWs) for all of them. Our selection
of lines and transition probabilities followed that of Lee \&
Carney (2002).

\section{Stellar Parameters}

Alonso et al. (2004) report on the analysis of 7 spectra of TrES-1
taken with the CfA Digital Speedometers (Latham 1992). By
comparing them with a library of synthetic spectra they derive
estimates of the effective temperature, surface gravity, and
metallicity as follows: $T_\mathrm{eff} = 5250 \pm 200$ K, $\log g
= 4.5\pm 0.5$, [Fe/H]$\sim 0.0$. As the authors later point out in
their discovery paper, these relatively large uncertainties have a
significant impact on the final estimates of the mass and radius
of the transiting planet, and ultimately on the possibility to
confront theory with observation. We describe below the four-fold
strategy we have devised to better constrain the stellar and
consequently planetary parameters, based on a detailed analysis of
our Keck and HET spectra.

\subsection{Atmospheric Parameters from Spectroscopy}

The stellar atmospheric parameters for TrES-1 were obtained using
a standard technique of Fe ionization balance (see for example
Santos et al. 2004, and references therein). The best-fit
configuration of the parameters is obtained by means of an
iterative procedure that encompasses the following three steps:
$a)$ the effective temperature is obtained by imposing that the
abundances $\log\varepsilon($\ion{Fe}{1}$)$ obtained from the
\ion{Fe}{1} lines be independent of the lower excitation potential
$\chi_l$ (i.e., zero correlation coefficient);  $b)$ the
microturbulent velocity $\xi_t$ is determined so that
$\log\varepsilon($\ion{Fe}{1}$)$ is independent of the reduced
equivalent widths $EW_\lambda/\lambda$; $c)$ the surface gravity
is determined by forcing exact agreement between abundances
derived from the \ion{Fe}{2} and those obtained from the
\ion{Fe}{1} lines. Uncertainties in the parameters were estimated
following the prescriptions of Neuforge \& Magain (1997) and
Gonzalez \& Vanture (1998), and rounded to 25 K in
$T_\mathrm{eff}$, 0.1 dex in $\log g$, and 0.05 km s$^{-1}$ in
$\xi_t$.

The final atmospheric parameters, along with the best-fit
\ion{Fe}{1} abundances, are: $T_\mathrm{eff} = 5250\pm 75$ K,
$\log g = 4.6\pm 0.2$, $\xi_t = 0.95\pm 0.1$ km s$^{-1}$, and
[Fe/H] $= 0.00\pm 0.09$. Including in the analysis EWs of 11
\ion{Fe}{1} and 4 \ion{Fe}{2} lines measured in the region of the
HRS averaged spectrum in common with the Keck template spectrum
gave identical results. The good agreement between the HIRES-only
and HIRES+HRS analysis suggests that significant sources of
systematic errors due to different spectrograph/configurations are
absent. Our spectroscopic estimates of $T_\mathrm{eff}$, $\log g$,
and [Fe/H] also agree remarkably well with the numbers reported by
Alonso et al. (2004).

Finally, a possible matter of concern are systematic
uncertainties inherent to the 1-D, LTE model atmospheres used in
the derivation of effective temperature and metallicity for a star
as cool as TrES-1.
However, recent studies (Yong et al. 2004) argue that non-LTE
effects arise primarily in stars significantly cooler than TrES-1,
and with metallicities significantly departing from solar (e.g.,
Asplund (2003), and references therein).

\subsection{$T_\mathrm{eff}$ Estimate from Photometry}

There is no direct distance estimate for TrES-1, and this prevents
the determination of an absolute luminosity. In addition, as the
star does not show signs of evolution off the main sequence, the
surface gravity constitutes a poor constraint for the models. The
effective temperature and metallicity, together with an age
estimate as we discuss below, are thus the only reliable metrics
that can be used in the comparison with theoretical isochrones. It
is then crucial to provide an independent estimate of
$T_\mathrm{eff}$ for an assessment of the validity of its
spectroscopic value and uncertainty.

To this end, we utilized the $BVJHK$ apparent magnitudes to derive
several $T_\mathrm{eff}$ estimates based on a number of empirical
color-temperature calibrations from the literature. The Johnson
$B$ and $V$ magnitudes for TrES-1 were obtained by means of
differential photometry (including differential extinction)
relative to a number of nearby stars (Mandushev, private
communication). For the apparent luminosity in the infrared
filters we relied upon the 2MASS catalog. From these values, and
their relative uncertainties, we derived empirical
$T_\mathrm{eff}$ estimates based on the color-temperature
calibrations of Mart{\'\i}nez-Roger et al. (1992), Alonso et al.
(1996), and Ram{\'\i}rez \& Mel\'endez (2004). We obtained an
average value for the effective temperature of $T_\mathrm{eff} =
5206\pm 92$ K. This result is in good agreement with our own
spectroscopic $T_\mathrm{eff}$ determinations reported in the
previous Section, giving us confidence in the spectroscopically
determined $T_\mathrm{eff}$ from Keck and HET.

\subsection{Stellar Activity and Age}

Inspection of the \ion{Ca}{2} H and K lines in the HIRES spectra
revealed a slight core reversal, indicating that TrES-1 is an
active star. In Figure~\ref{activity1} we show a region of the
HIRES template spectrum centered on the \ion{Ca}{2} H line. For
comparison, we have overplotted the spectrum of an old, inactive
star (HIP 86830) with the same temperature from the Sozzetti et
al. (2004) sample of metal-poor stars, which has a metallicity 
[Fe/H] $= -0.68$. The emission feature is clearly evident. A
possible explanation for the activity level is that the star is
not very old. To provide useful constraints on its age, we have
collected three different pieces of external evidence.

First, we have measured the chromospheric activity index S from
the Ca H and K lines, based on the prescriptions by Duncan et al.
(1991), for TrES-1 and for two Hyades stars (HD 28462 and HD
32347) of the same spectral type observed in the context of the
G-dwarf Planet Search program (Latham 2000). We transformed our S
values to the standard Mount Wilson S index applying the relation
derived by Paulson et al. (2002). We then applied relations from
Noyes et al. (1984, and references therein) to convert to the
chromospheric emission ratio $R^\prime_\mathrm{HK}$ and to derive
estimates of the age $t$ for the three stars. We obtained $\log
R^\prime_\mathrm{HK} = -4.77$ for TrES-1. The activity levels for
the two Hyades stars are in excellent agreement with the values
reported by Paulson et al. (2002), giving us confidence that our
$R^\prime_\mathrm{HK}$ for TrES-1 is reliable. As its activity
level is about half that of the two Hyades members, but higher
than that reported for its very close analog $\alpha$ Cen B ($\log
R^\prime_\mathrm{HK} = -4.92$; Chmielewski 2000), we can argue
that TrES-1 is older than Hyades stars of the same spectral type,
but younger than $\alpha$ Cen B ($\sim$4.2 Gyr; Henry et al.
1996). Indeed, based on the Noyes et al. (1984) relations, we find
an age value $t = 2.5$ Gyr. We note, however, that our measure of
$R^\prime_\mathrm{HK}$ is based on a single-epoch observation, thus 
additional \ion{Ca}{2} meausurements are clearly encouraged. 

Secondly, following Gonzalez (1998), using the atmospheric
parameters derived from the Fe-line analysis, we synthesized a 10
\AA\, region of the spectrum including the 6707.8 \AA\, Li line in
the HRS averaged spectrum. In Figure~\ref{activity2} we show the
comparison of the spectrum of TrES-1 with three models, each
differing only in the Li abundance assumed. The lithium line is
not detectable by eye in the noise. We place an upper
limit for the Li abundance of $\log\varepsilon(\mathrm{Li}) <
0.1$, consistent with the star being older than a Hyades star of
the same temperature.

Finally, due to a fast decrease in activity levels for ages
greater than $\sim 1$ Gyr (Pace \& Pasquini 2004), at $\sim 2$ Gyr
the emission in \ion{Ca}{2} is down by a factor of a few with
respect to the Hyades, and the value of $R^\prime_\mathrm{HK}$ for
TrES-1 is consistent with this interpretation. Our preliminary age
estimate for TrES-1 is then $2.5\pm 1.5$ Gyr.

\subsection{Stellar Mass and Radius}

Using the constraints described above on $T_\mathrm{eff}$, [Fe/H],
and the age of the star, we used stellar evolution models and ran
Monte Carlo simulations to infer the stellar mass and radius for
TrES-1 as well as their uncertainties. The simulations were
repeated using two different evolutionary models: Girardi et al.
(2000) and Yi et al. (2003). We obtain very good agreement in the
derived parameters in both cases, to within 0.01 solar units in
both mass and radius, despite slightly different assumptions in
the input physics of these two sets of models.
We report here the results for TrES-1 using the Girardi et al.
(2000) models: $M_\star = 0.89\pm 0.05$ $M_\odot$, $R_\star =
0.83\pm 0.05$ $R_\odot$. These error bars include a somewhat
arbitrary contribution of 0.04 added in quadrature to the formal
Monte Carlo errors of 0.03 in both mass and radius, to account for
unforeseen systematics in both the stellar evolution and stellar
atmosphere models, as well as other physical assumptions. The
uncertainties on metallicity and effective temperature contribute
about equally to the errors on both mass and radius, while the age
uncertainty contributes negligibly.

Finally, based on our simulations, we estimate an absolute
magnitude $M_v = 5.85\pm 0.15$, and thus place TrES-1 at a nominal
distance $d \simeq 150$ pc. Its resulting galactic velocity vector
is [$U$,$V$,$W$] = [-15, -33, 16] km s$^{-1}$.

\section{Planetary Parameters}

Following Alonso et al. (2004), we utilized the revised values of
$M_\star$ and $R_\star$, and their uncertainties, in a new
$\chi^2$ minimization procedure of the photometric points for the
model light curve as a function of $R_p$ and $i$. We find $R_p =
1.04^{+0.08}_{-0.05}$ $R_\mathrm{J}$, and $i =
89^\circ.5^{+0.5}_{-1.3}$. The values and uncertainties for the
primary mass, inclination, and the radial velocity semi-amplitude
are then combined to give a value of the planetary mass $M_p =
0.76\pm 0.05$ $M_\mathrm{J}$. With the improved value of the
stellar mass, now the dominant contribution to the planet mass
uncertainty comes from the spectroscopic orbit, while the
contribution due to the error in the inclination angle is
negligible.

While the revised value for $M_p$ is essentially identical to the
one reported by Alonso et al. (2004), the planetary radius found
here is slightly smaller, due to the smaller stellar radius
estimate obtained in our analysis.

\section{Summary and Discussion}

We have derived new values of the stellar atmospheric parameters
of the transiting planet host-star TrES-1 from high-resolution,
high S/N spectra. Our spectroscopic $T_\mathrm{eff}$ values are in
very good agreement with empirical color-temperature calibrations
and with the estimates reported by Alonso et al. (2004). The star
is a main-sequence object with a metallicity consistent with
solar. The lack of detectable lithium argues both against youth
and recent pollution events due to ingestion of planetary
material. Its measured \ion{Ca}{2} activity levels are further
suggestive of an object with an age of a few Gyr, an intermediate
value between the Hyades and $\alpha$ Cen B.

Based on the new constraints on atmospheric parameters and age, we
have significantly improved on the determination of the stellar
and planetary mass and radius. Our updated values for $M_\star$
and $R_\star$ are consistent with those reported by Alonso et al.
(2004), although we find a slightly smaller radius and a slightly
larger mass. The most significant improvement comes at the level
of the determination of $R_p$ and $M_p$ with our new analysis of
the transit photometry. In Figure~\ref{massrad} we show masses and
radii for all transiting planets known to-date, including TrES-1.
Thanks to the more accurate values of the stellar parameters, the
radius and mass of the planet have reduced uncertainties with
respect to those reported by Alonso et al. (2004). In particular,
for very similar values of $M_p$, the difference between the
values of $R_p$ for HD 209458 and TrES-1, about $25 \%$, is now
significant at the 3-$\sigma$ level.

The well-determined radius of TrES-1 opens the door to new
important tests of theoretical evolutionary models of irradiated
extrasolar giant planets (e.g., Bodenheimer et al. 2003; Burrows
et al. 2004; Chabrier et al. 2004), as its parent star is about
1000 K cooler than HD 209458 and OGLE-TR-56, the other two objects
that have recently been targets of extensive investigations. Its
radius seems to agree better with the predictions from models that
do not invoke additional heat/power sources in the core (Showman
\& Guillot 2002; Baraffe et al. 2003), or tidal heating effects
due to the gravitational perturbation of an undetected long-period
companion (Bodenheimer et al. 2003). Given its orbital radius and
the characteristics of the host star, its mass loss rate (Baraffe
et al. 2004) is also likely to be significantly reduced with
respect to those, for example, expected for OGLE-TR-56b and
OGLE-TR-132b.

Additional detections of giant planets transiting relatively
bright stars covering a range of spectral types are clearly
necessary for a continued improvement of our understanding of the
structure and evolutionary properties of this peculiar class of
objects.

\acknowledgments

Over the course of this investigation, we have benefited from
stimulating discussion with various colleagues, in particular G.
Mandushev, R. Alonso, H. Deeg, J. Belmonte, E. Dunham, and F.
O'Donovan. We are thankful to the Director of the McDonald
Observatory, D. Lambert, for providing Director's discretionary
time, and to the HET staff, in particular S. Odewahn, V. Riley,
and M. Villareal for their assistance. The Hobby-Eberly Telescope
is operated by the McDonald Observatory on behalf of the
University of Texas at Austin, the Pennsylvania State University,
Stanford University, Ludwig Maximillians Universit\"at M\"unchen,
and Georg August Universit\"at G\"ottingen. Some of the data
presented herein were obtained at the W.M. Keck Observatory, which
is operated as a scientific partnership among the California
Institute of Technology, the University of California and the
National Aeronautics and Space Administration.

\clearpage

\figcaption{Comparison between the \ion{Ca}{2} H line for TrES-1
(solid line) and an inactive star of the same temperature (dotted
line)\label{activity1} }

\figcaption{A portion (upper panel) of the HET averaged spectrum
of TrES-1 containing the \ion{Li}{1} line at 6707.8 \AA\, (filled
dots), compared to three syntheses (lines of various colors and
styles), each differing only for the lithium abundance assumed.
The bottom panel shows the data on an enlarged scale, so the model
spectra for low Li abundance can be
distinguished\label{activity2}}

\figcaption{Radius vs. mass for the sample of transiting
extrasolar planets discovered to-date, including TrES-1 (this
work). The values and uncertainties are taken from Brown et al.
(2001) for HD 209458b, Torres et al. (2004) for OGLE-TR-56b,
Bouchy et al. (2004) for OGLE-TR-113b, Moutou et al. (2004) for
OGLE-TR-113b, and Pont et al. (2004) for
OGLE-TR-111b\label{massrad} }


\begin{figure}
\plotone{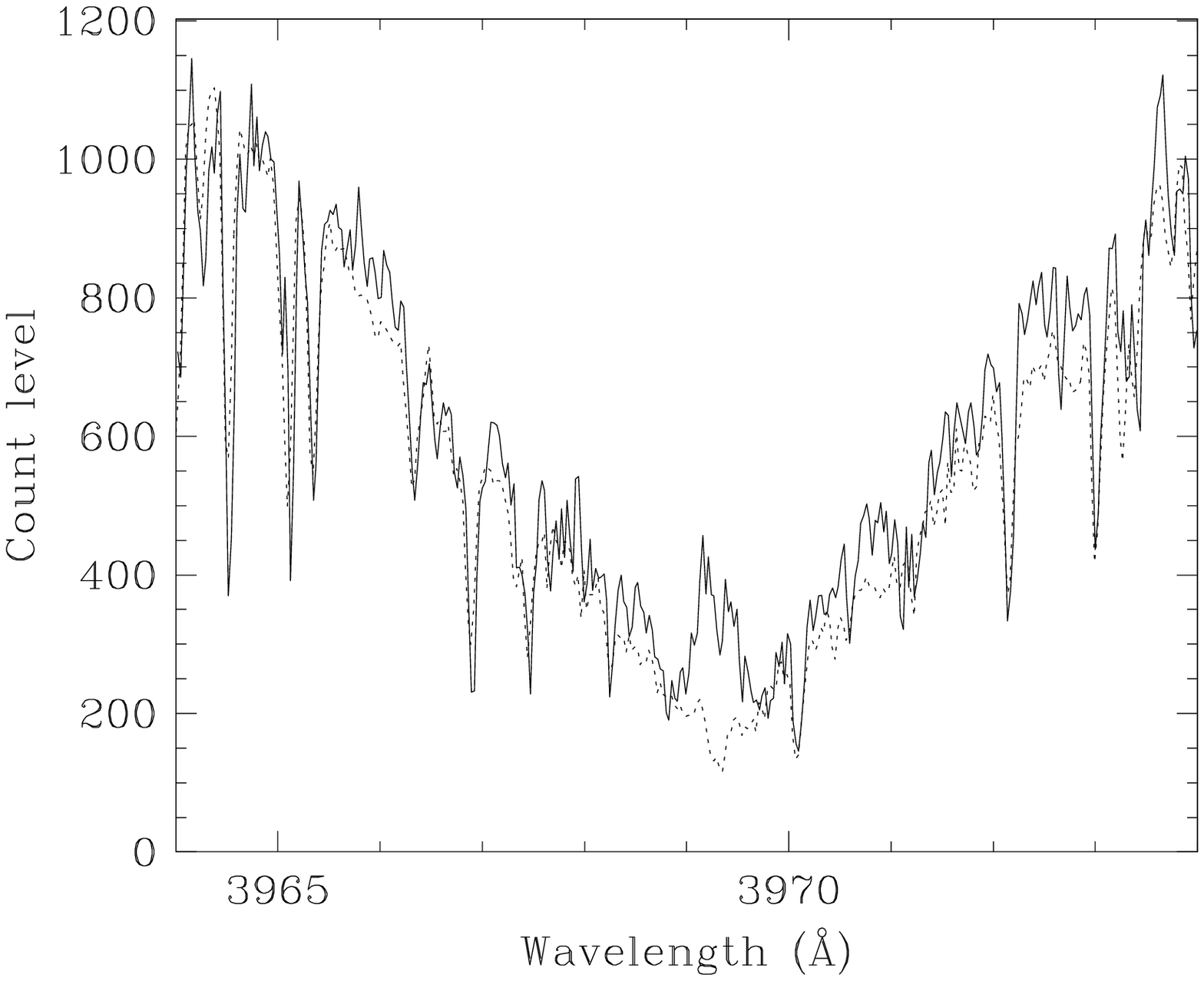}
\end{figure}

\clearpage

\begin{figure}
\plotone{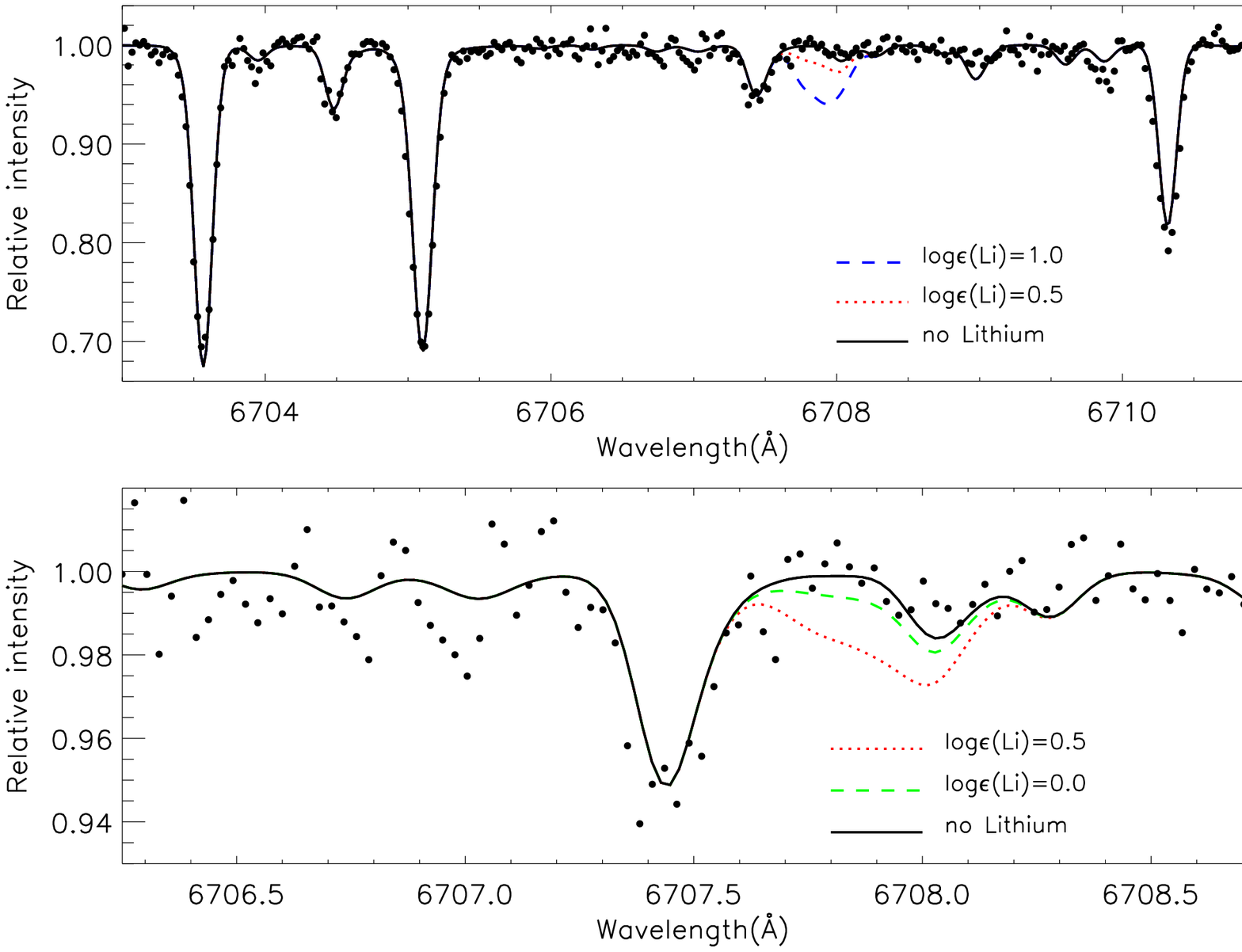}
\end{figure}

\clearpage

\begin{figure}
\plotone{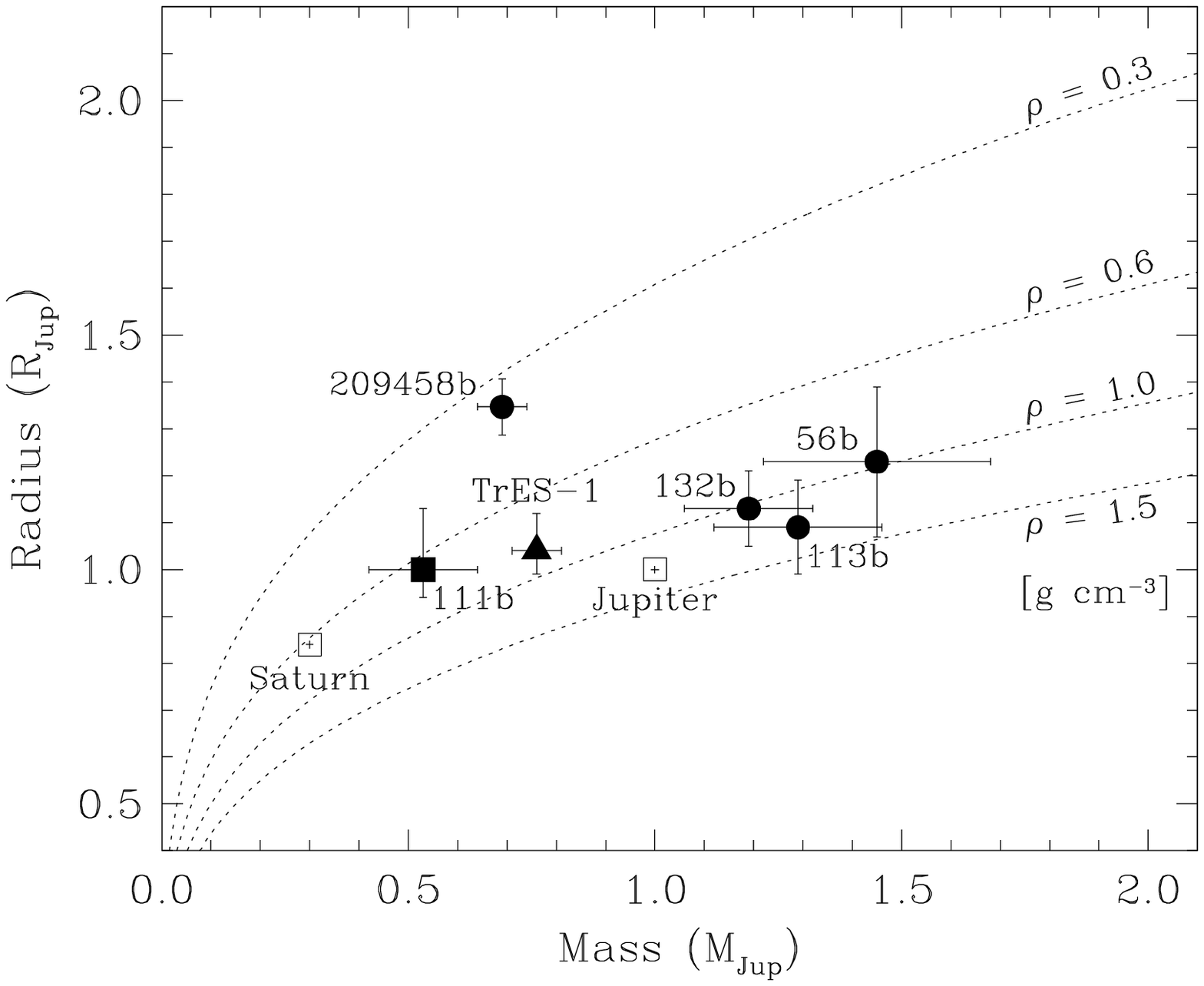}
\end{figure}

\end{document}